# Electron counting and a large family of two-dimensional semiconductors

*Mao-sheng Miao[1,2,3,*], Jorge Botana[3], Jingyao Liu[4] and Wen Yang[3]*

[1] Department of Chemistry and Biochemistry, California State University Northridge, CA 91330, USA

[2] Materials Research Laboratory, University of California, Santa Barbara, CA 93106-5050, USA

[3] Beijing Computational Science Research Center, Beijing 10084, P. R. China

[4] *Institute of Theoretical Chemistry, Jilin University, Changchun, P. R. China*

* Email: maoshm@ucsb.edu

**Abstract**

Comparing with the conventional semiconductors, the choice of the two dimensional semiconductor (2DSC) materials is very limited. Based on proper electron counting, we propose a large family of 2DSCs, all adopting the same structure and consisting of only main group elements. Using advanced density functional calculations, we demonstrate the attainability of these materials, and show that they cover a large range of lattice constants, band gaps and band edge states, therefore are good candidate materials for heterojunctions. This family of two dimensional materials may pave a way toward fabrication of 2DSC devices at the same thriving level as 3D semiconductors.

**Keywords**

Novel two-dimensional semiconductors; large band gap variation; two-dimensional heterojunctions; density functional theory; hybrid functional

Two-dimensional semiconductors (2DSC) are currently the focus of many studies in condensed matter and materials research, thanks to their novel and superior transport properties that may shape the future electronic devices.[1] They have been intensively studied in the areas of low-dimension electronics,[2,3] topological insulators[4] and valleytronics[5,6] as well as solar energy harvesting such as photolysis.[7] Despite the high expectation for future electronics,[1,2,4-6,8] there are only a few 2D semiconductors (2DSCs) materials available, which adopt very different atomic structures.[9,10] Besides a couple of light element materials including graphene, h-BN, g-$C_3N_4$, the other options are the transition metal dichalcogenides ($MX_2$).[11-13] However, most of the $MX_2$ compounds are metallic and also magnetic. Only a few 4d and 5d $MX_2$ are semiconducting, covering a small range of gap variation ( 0.5 – 1.6 eV) and a small lattice variation from about 3.2 Å to 3.5 Å.[14]

In contrast, the prosperity of the information technology in the last several decades greatly depends on the rich choice of 3D semiconductor (3DSC) materials.[15] These materials consist primarily of the main group elements.[16] The advantage of main group compounds is that the atoms have fixed valence, which ensures that the correct electron counting can lead to compounds with a gap. It also provides a clear route to n-type and p-type doping by replacing the elements with excess or deficient electrons in the valence orbitals.[16] The band gaps of the 3DSCs cover a large energy spectrum from infrared (< 1.5 eV) to ultraviolet (> 3.3 eV). Despite the large variety of the properties, most of the 3DSCs adopt the same diamond (elementary) or zinc blend (binary compounds) structures, with several exceptions adopting the similar wurtzite (WZ) structure.

In 3DSC structures, atoms tetrahedrally bonded with their neighbors. For example, Si and Ge are in the diamond structure in which each atom bonded with four neighboring atoms. Since each $sp^3$ bond contribute one electron, Si and Ge atoms form four σ bonds and each atom satisfies the octet rule. Similarly, group III elements such as Ga will form semiconductors with group V elements such as As, in which group III and V atoms contribute 3 and 5 electrons respectively. II-IV semiconductors follow a similar trend. In contrast, Mo in $MoS_2$ has a valence of 4 and forms 6 covalent bonds with the surrounding S atoms (Fig. 1a). Each S atom forms 3 bonds with the neighboring Mo atoms and contains one $sp^3$ lone pair of electrons pointing out of the $MoS_2$ plane. If replacing each Mo atom by two covalently bonded group III elements such as Ga (Fig. 1b), the electron counting will be the same as $MoS_2$. In the corresponding $Ga_2S_2$ or GaS compounds, the electrons will occupy all the bonding states and leave all anti-bonding states empty, forming a semiconductor. Different to $MoS_2$ that greatly favors the Inversion Asymmetric (IaS) structure, these III-VI 2DSCs can have both IaS and Inversion Symmetric (IS) structures (Fig. 1b and 1c). Extending the structure to other main group elements, we find that group IV elements can form similar 2DSCs with group V elements, group II elements can form 2DSCs with group VII elements (Fig. 1d).

Our calculations are performed using plane-wave based density functional theory (DFT) method as implemented in the Vienna Atomic Simulation Package (VASP).[17] The ionic potentials are described by the project augmented wave (PAW) method.[18] The geometry and the electronic structures are relaxed by both a semi-local functional in the framework of Purdew-Burke-Ernzerhof (PBE)[19] and a hybrid functional in the framework of Heyd-Scuseria-Ernzerhof (HSE).[20] The results shown in the figures are obtained by HSE functional. Comparing with PBE, the HSE can largely improve the band gaps for 3D semiconductors and insulators. Its changes to the lattice constants of the materials are usually quite small.

Using DFT calculations, we first examine the stability of the proposed 2DSC compounds. Assuming A is cation and B is anion, we compare the energy of single layer AB and the energy of the most stable AB (1:1) compound. In case there is no stable 1:1 AB compound, we calculate the formation energy of AB from reacting the most stable $A_nB_m$ compound with necessary excessive A or B elementary solid. If there is no known stable A-B compound, we calculate the formation energy of AB 2DSC from reacting the A and B elementary solids. The results are summarized in Fig.2. As shown, many IV-V and III-VI 2DSCs are stable compared with their 3D compounds. For example, 2D SiP is about 23 meV lower in energy than its 3D counter part. For those that are not the most stable, many are energetically close to their most stable form. We would like to emphasize that the above stability assessment is not conclusive, since our calculations cannot cover all the possible compositions and structures for all the compounds. However, it is well known that two-dimensional structures can be stabilized and fabricated with the help of substrates. A well-known example is silicene that is significantly higher in energy than bulk silicon and yet can be obtained on certain substrates.[21,22] Therefore, it is reasonable to expect that the proposed 2DSCs attainable, as supported by the energy comparisons with known compounds.

Another interesting feature of the 2DSCs, as revealed in Fig. 2, is that the energy differences between the IaS and the IS structures are quite small. For most of the 2DSCs except CN and CP, the IaS structures are more stable. The energy difference varies but is generally smaller than 35 meV, in contrast to the large energy difference of 400 meV for $MoS_2$. For both AB and $MX_2$, the difference between IaS and IS structures is the relative positioning of the lower and upper anion layers. However, the low rotational barrier of the A-A bonds in AB 2DSC determines that the two polytype structures are close in energy. On the other hand, the $MX_6$ octahedral is the result of the transition metal $sp^3d^2$ hybridization and is very directional. That is why their IaS and the IS structures have large energy difference.

Many IV-V, III-VI and II-VII combinations can form 1:1 stoichiometric 3D compounds, many with structures that further indicates the possibility of fabricating the corresponding 2DSCs. For example, SiP forms a Cmc21 structure (Fig. 3a) which consists of SiP layers. In each layer, the local Si-Si and Si-P bonding is identical to SiP 2DSC. Different to 2DSC, some of the $Si_2P_6$ groups point to in-plane directions. Si-P can also form a $SiP_2$ compound that is stable in the pyrite structure (Fig. 3b). GeP, GeAs and GeSb can be found in an I4mm structure in which each Ge bonds with five neighboring anions, and vise versa. However, as shown in Fig. 2, these structures are less stable than the 2DSC layered structure. The Sn-V compounds are either in I4mm structure (SnP) or in NaCl structure (SnAs and SnSb). SnBi forms alloy instead of stoichiometric compound. Pb and group V elements do not form stable compounds of any kind. Our calculations show that these 2D layered structures are not stable. However, the interaction with the proper substrate surfaces may stabilize them.

There is not any known Aluminum-VI compound. However, our calculations show that the Al-VI 2DSCs are very stable against the decomposition into elements. GeS, GeSe and GeTe are known to form layered structure (P63/mmc, Fig. 3d). It is interesting that single layer Ge-VI has been successfully fabricated through mechanical exfoliation of the layered 3D compounds.[23-26] InS is stable in an orthorhombic Pmnn structure (Fig. 3e) which also consists of the $In_2S_6$ structural units that form a 3D netweork. [27] Our calculations show that the InS 2DSC is slightly more stable than the Pmnn InS. InSe exhibit the same 3D layered structure as GaSe (P63/mmc). InTe is found in $Tl_2Se_2$ structure (I4/mcm, Fig. 3f), which is higher in energy than the 2DSC InTe. Hg-VII can form both 1:1 and 1:2 stoichiometric compounds. HgBr and HgI are both found in a HgCl structure which can be viewed as a layered array of Hg-VII molecules (Fig. 2i). $HgI_2$ and $HgBr_2$ adopt $P4_2/nmc$ and $Cmc2_1$ structure respectively (Fig. 2g and 2h). The P42/nmc is a layered structure; however its building units are the Hg-$I_4$ tetrahedrons. This structure is slightly more stable than the proposed HgI 2DSC structure.

While adopting the same structure, 2DSCs cover a large range of energy gaps as well as the lattice constants. As shown in Fig. 4a, the energy gap ranges from 5.26 eV for CN or 3.29 eV for GaS to values close to or below 1eV for GeSb, Pb compounds and Bi compounds. The lattice constants also vary from 2.35 Å for CN to 4.65 Å for PbBi. This large variation of the energy gaps and lattice constants may allow the fabrication of various heterojunctions for electronic and optoelectronic devices. For comparison, we also show the band gaps and lattice constants of $MoX_2$ and $WX_2$ (X=S, Se and Te). As shown in Fig. 4a, this series of compounds cover the energy gap range from about 2.02 eV ($MoS_2$) to 1.03 eV ($WTe_2$). The lattice constants only range from 3.16 to 3.52 Å.

Furthermore, comparing with 3DSCs, the 2DSCs are usually more sustainable to very large strains that can significantly modify their electronic structures.[28] For example, $MoS_2$ can maintain the structure under strains as large as 11%,[29,30]and strains as large as 18.5% has been found for bended graphene.[31] The large strains can be used to approach a narrower band gap. As shown in Fig S2 in the supplementary information, the band gap of SnSb can be reduced from 1.40 eV to 0.31 eV under a strain of 13.6%.

For 3D semiconductors, the band edge states can be calculated by a slab model that contains a region of semiconductor and a region of vacuum.[32] The VBM and the CBM positions relative to vacuum level can be obtained by comparing the electrostatic potentials at the centers of semiconductor and vacuum regions. For 2DSC calculations, there is already a large vacuum region. The edge states are calculated as following: $E_{\mathrm{VBM}} = E_{\mathrm{VBM}}^{calc} - \left(\Phi_{\mathrm{avg}} - \Phi_{\mathrm{vac}}\right)$, in which $E_{\mathrm{VBM}}^{calc}$ is the calculated VBM, $\Phi_{\mathrm{avg}}$ and $\Phi_{\mathrm{vac}}$ are the average electrostatic potential and the electrostatic potential at the center of the vacuum region (Fig. S3).

Fig. 4b shows the band-edge positions of the 2DSCs relative to the vacuum level. The proposed family of 2DSC not only has large ranges of band gaps and lattice constants, their band edges, including the valence band maximum (VBM) and the conduction band minimum (CBM), are also very different. The variation is especially significant for IV-V compounds. The general trend of the band edge states is similar to 3DSC, *i.e.* the VBM increases (or ionization energy decreases) with increasing atomic number of the anions. This is due to the fact that VBM states consist of mainly the anion p states and become higher in energy with increasing atomic number in the same group. Furthermore, although the IS and IaS are very close in fundamental band gaps, for many compounds especially IV-Vs, they show noticeable difference in the band edge states.

In conclusion, we proposed a large family of two-dimensional semiconductor materials in an identical structure that consist of only main group elements and satisfy the electron counting. Using density functional method with hybrid functionals, we show that the band gap and the band edge states of the proposed materials cover a large range of variation. By comparing the energies with the existing compounds, we also demonstrate that the majority of the materials are attainable, especially with the help of proper choice of substrates and surfaces that has been proven to be an effective approach to stabilize single layer materials. Obtaining these materials may advance the 2D electronics to a completely new level.

## Figure Legends

**Figure 1.** Schematics of main group 2D semiconductor materials. (A) Structure of single layer $MoS_2$; (B) Structure of Inversion Asymmetric 2DSC; (C) Structure of Inversion Symmetric 2DSC; (D) Schematics of possible combinations of elements in the periodic table that may form three dimensional semiconductors and 2DSCs.

**Figure 2.** Stability of the 2D main group semiconductors. The calculated energy difference between layered 2DSCs and the selected compounds with corresponding compositions. The bars show the energy differences between the IS and the IaS structures. For all the compounds except CN and CP (marked as orange), the IaS structure is lower in energy than the IS structure.

**Figure 3.** Structures of the related 2D and 3D materials. (A) Quasi-layered structure of SiP ($Cmc2_1$); (B) $SiP_2$ in pyrite structure; (C) I4mm structure for GeP, GeAs, and GeSb; (D) GaS layered structure ($P6_3/mmc$); (E) InS structure (Pmnn); (F) InTe in $Tl_2Se_2$ structure (I4/mcm); (G) $HgI_2$ structure ($P4_2/nmc$); (H) $HgBr_2$ structure ($Cmc2_1$); (I) HgBr in HgCl structure (I4/mmm).

**Figure 4.** Electronic structures of the 2D main group semiconductor materials. (A) Distribution of band gaps and lattice constants of proposed 2DSCs. (B) The band edge states of 2DSCs. The blue and the orange bars represent the 2DSCs in IaS and IS structures, respectively. The upper and lower edges of the bars show the CBM and the VBM positions relative to the vacuum.

**Acknowledgement** M. S. M is partially supported by the MRSEC program (NSF-DMR1121053) and the ConvEne-IGERT Program (NSF-DGE 0801627). W. Y. is supported by National Science Foundation of China (Contract Nos. NSFC 11274036 and NSFC 11322542) and the Ministry of Science and Technology of China (Contract No. MOST 2014CB848701). Calculations are performed on NSF-funded XSEDE resources (TG-DMR130005), on resources in Center for Scientific Computing supported by the CNSI, MRL and NSF CNS-0960316, and on Beijing CSRC computing resources.

**Supporting Information**. Contains three additional figures, showing the bandgap coverage of 3D semiconductors, the bandgap engineering of 2D semiconductors through strains, and the alignment of the electrostatic potentials.

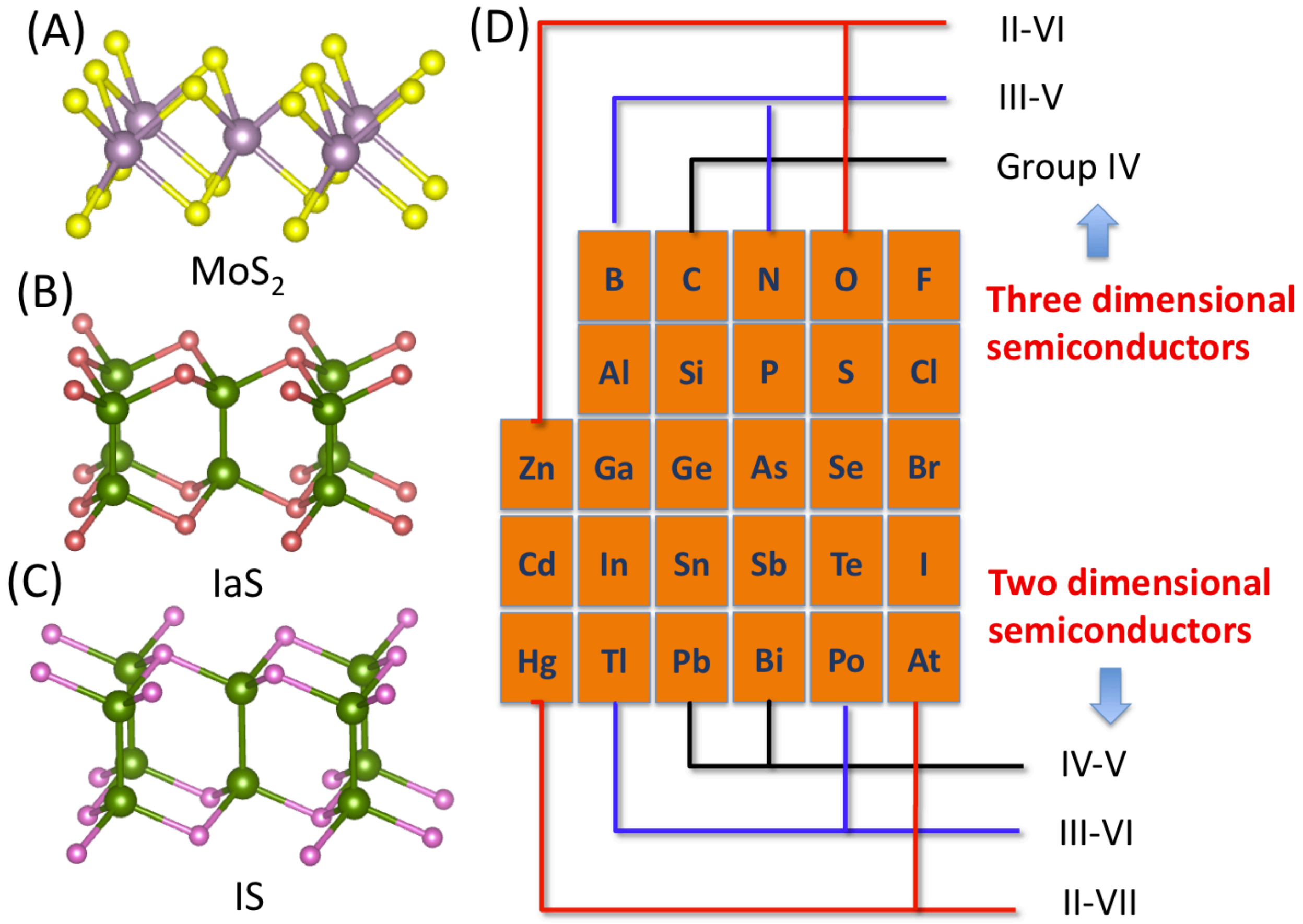

**Figure 1.**

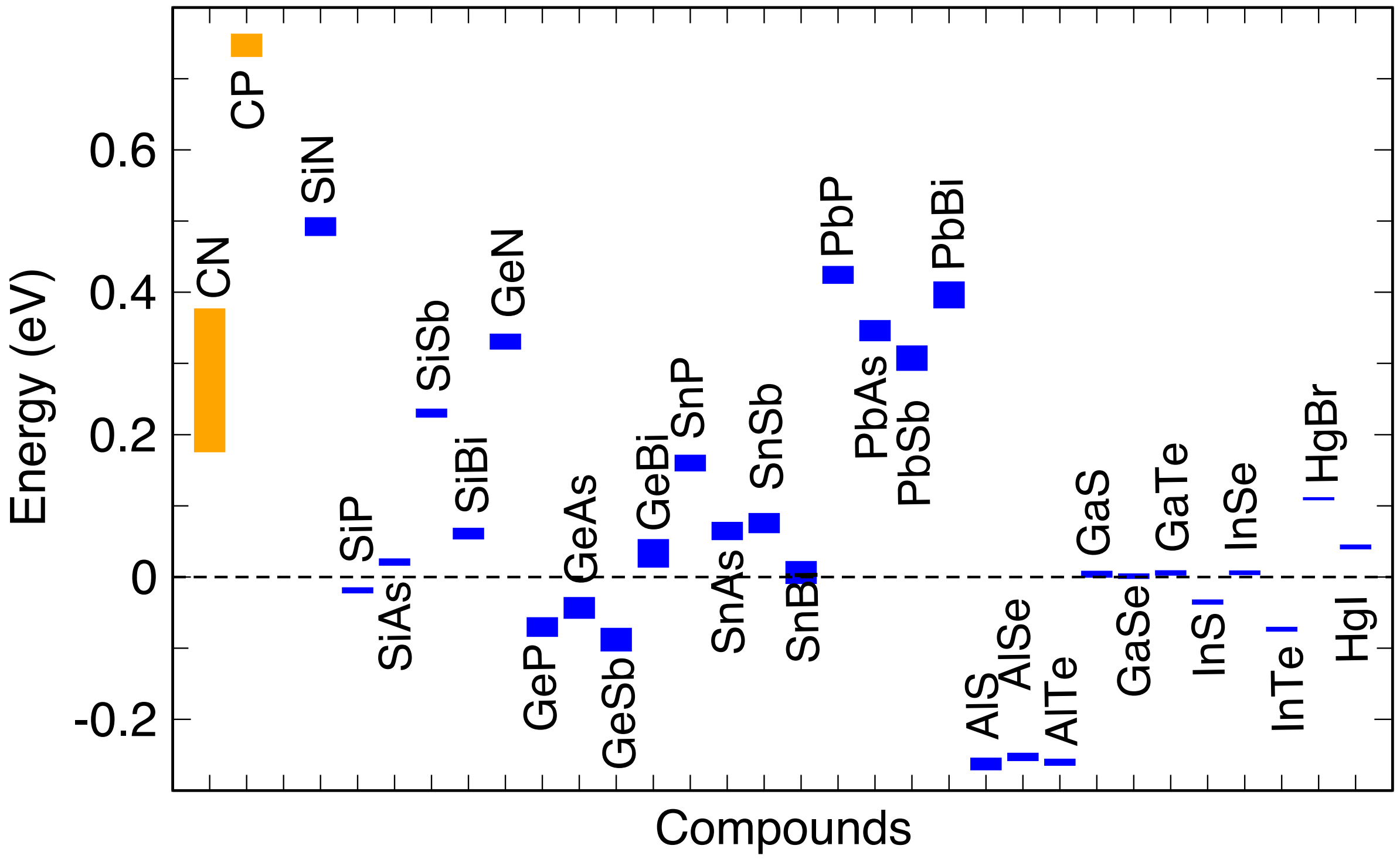

**Figure 2.**

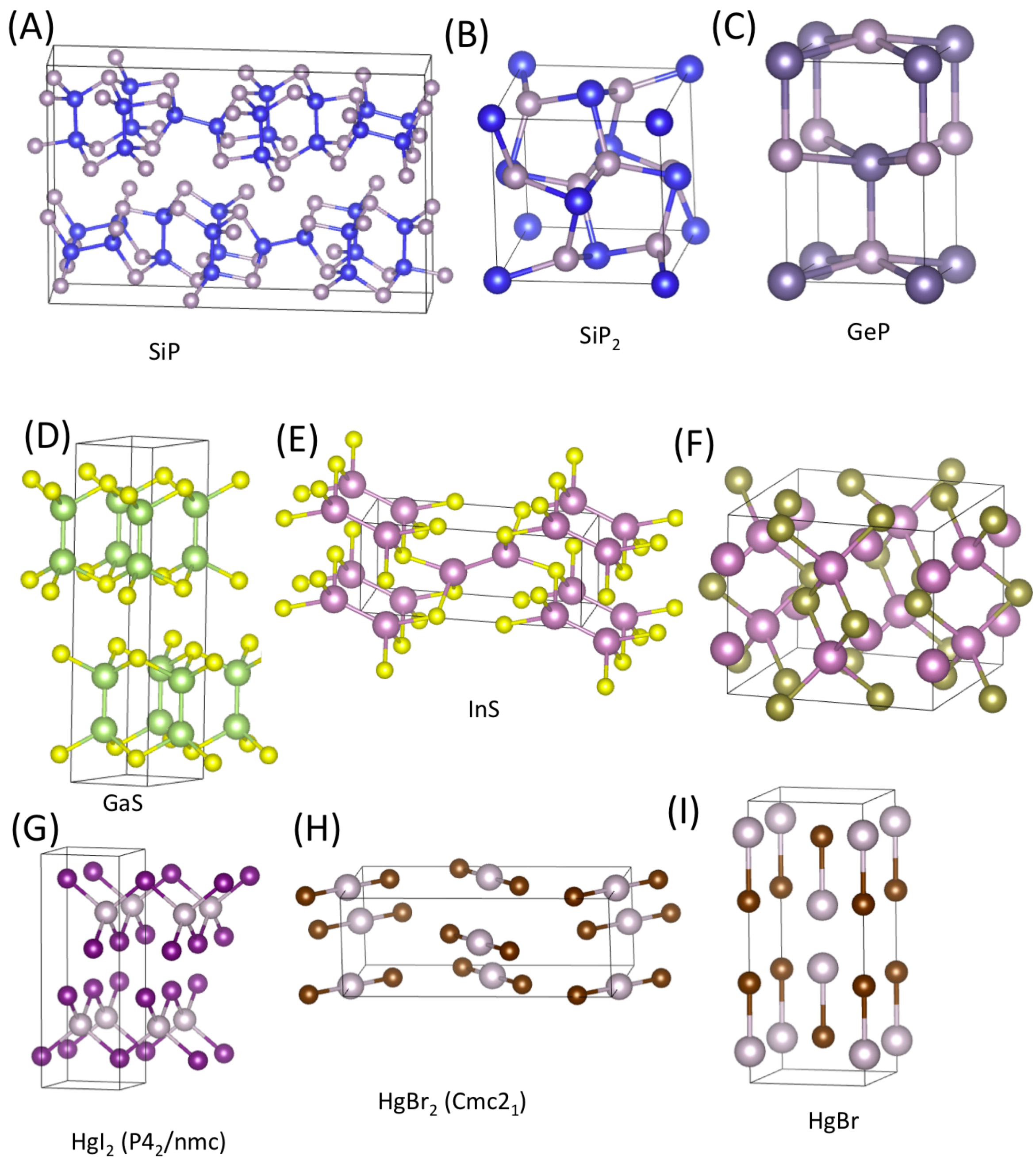

**Figure 3.**

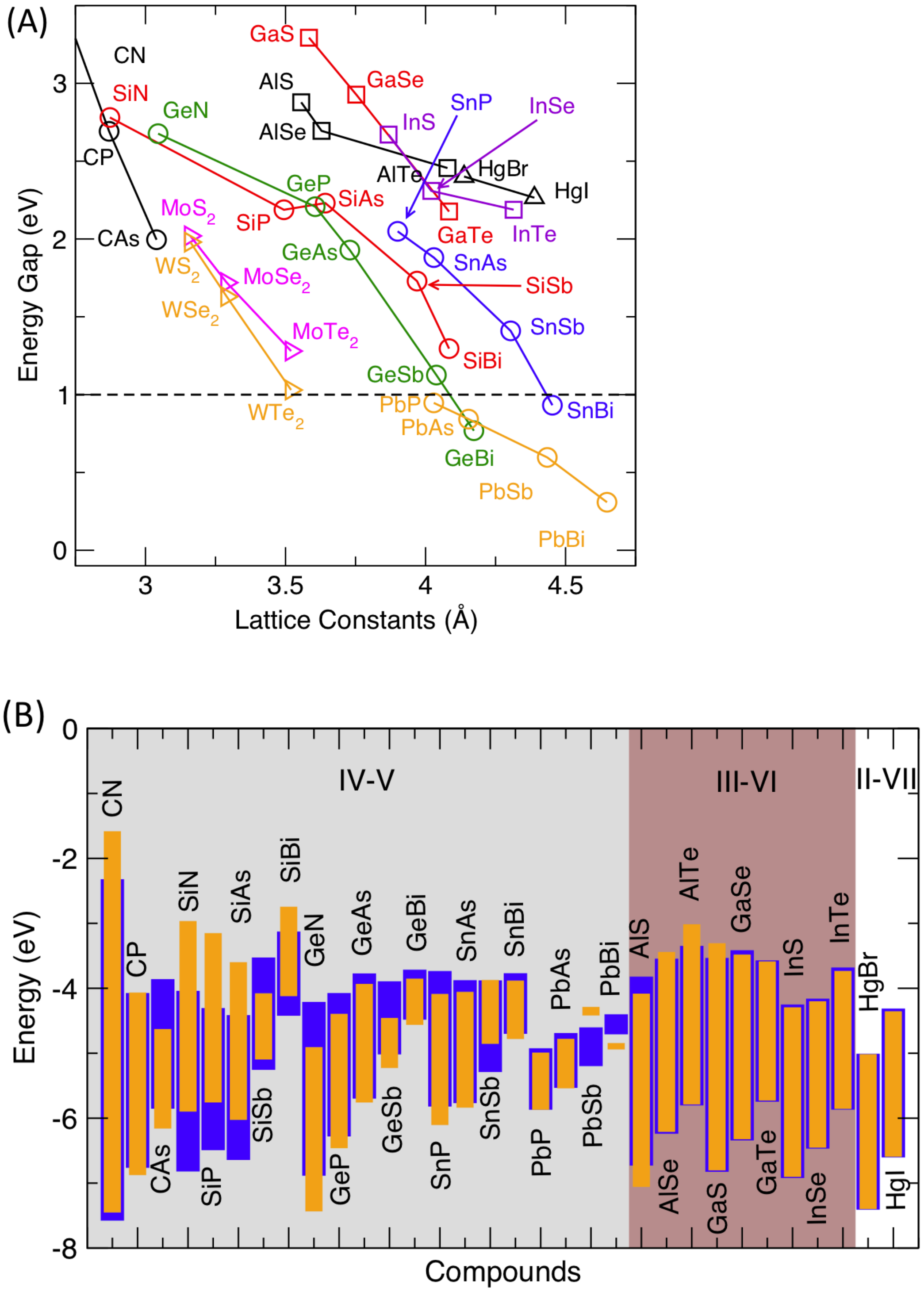

**Figure 4.**